\documentclass{iaus}
\usepackage{graphicx}

\title[Differential rotation from dynamo models] 
{Radial differential rotation vs surface differential rotation: investigation 
based on dynamo models}

\author[Korhonen \& Elstner]   
{H. Korhonen$^1$ \and D. Elstner$^2$}

\affiliation{$^1$ ESO, Karl-Schwarzschild-Str. 2, D-85748 Garching bei 
M{\"u}nchen, Germany\\
[\affilskip]
$^2$Astrophysical Institute Potsdam, An der Sternwarte 16, D-14482 Potsdam, 
Germany}

\pubyear{2009}
\volume{IAU259}  
\pagerange{119--126}
\date{?? and in revised form ??}
\jname{Proceedings Title IAU Symposium}
\editors{A.C. Editor, B.D. Editor \& C.E. Editor, eds.}
\begin{document}

\maketitle

\begin{abstract}
Differential rotation plays a crucial role in the alpha-omega dynamo,
and thus also in creation of magnetic fields in stars with convective outer
envelopes. Still, measuring the radial differential rotation on stars is
impossible with the current techniques, and even the measurement of surface
differential rotation is difficult. In this work we investigate the surface
differential rotation obtained from dynamo models using similar techniques as
are used on observations, and compare the results with the known radial
differential rotation used when creating the Dynamo model.
\end{abstract}


\section{Model and analysis methods}

The model used in this work was originally used to investigate the flip-flop 
phenomenon (\cite{elst_kor}, \cite{kor_elst}). In the current work is is used 
to study the correlations between surface and radial differential rotation. The
dynamo is modelled with a turbulent fluid in a spherical shell. The internal 
rotation is similar to the solar one with equatorial regions rotating faster 
than the polar regions, but here a smaller difference between core and surface 
rotation is used. The inner boundary of the convective zone is at the radius 
0.4R$\star$. The mean electromotive force contains an anisotropic alpha-effect 
and a turbulent diffusivity. The nonlinear feedback of the magnetic field acts 
on the turbulence only. The boundary conditions describe a perfect conducting 
fluid at the bottom of the convection zone and at the stellar surface the 
magnetic field matches the vacuum field. 

\begin{figure}
  \begin{center}
  \includegraphics[width=12.5cm]{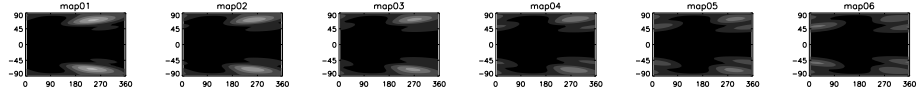}
  \caption{Six examples of the 36 snapshots of the Dynamo model.}
  \label{fig:maps}
  \end{center}
\end{figure}

We use standard cross-correlation methods to study the changes in the magnetic 
pressure maps obtained from the dynamo calculations. As these maps can be 
treated the same way as the temperature maps obtained from Doppler imaging, we 
can use the same techniques as in the case of the real observations to 
analyse the snapshots from the dynamo calculations. We have taken magnetic 
pressure maps from 36 time points over the activity cycle. The chosen time 
points, which are separated by about 50 days, are shown in Fig.~\ref{fig:maps}.

\section{Results}

\begin{figure}
  \begin{center}
  \includegraphics[width=12.5cm]{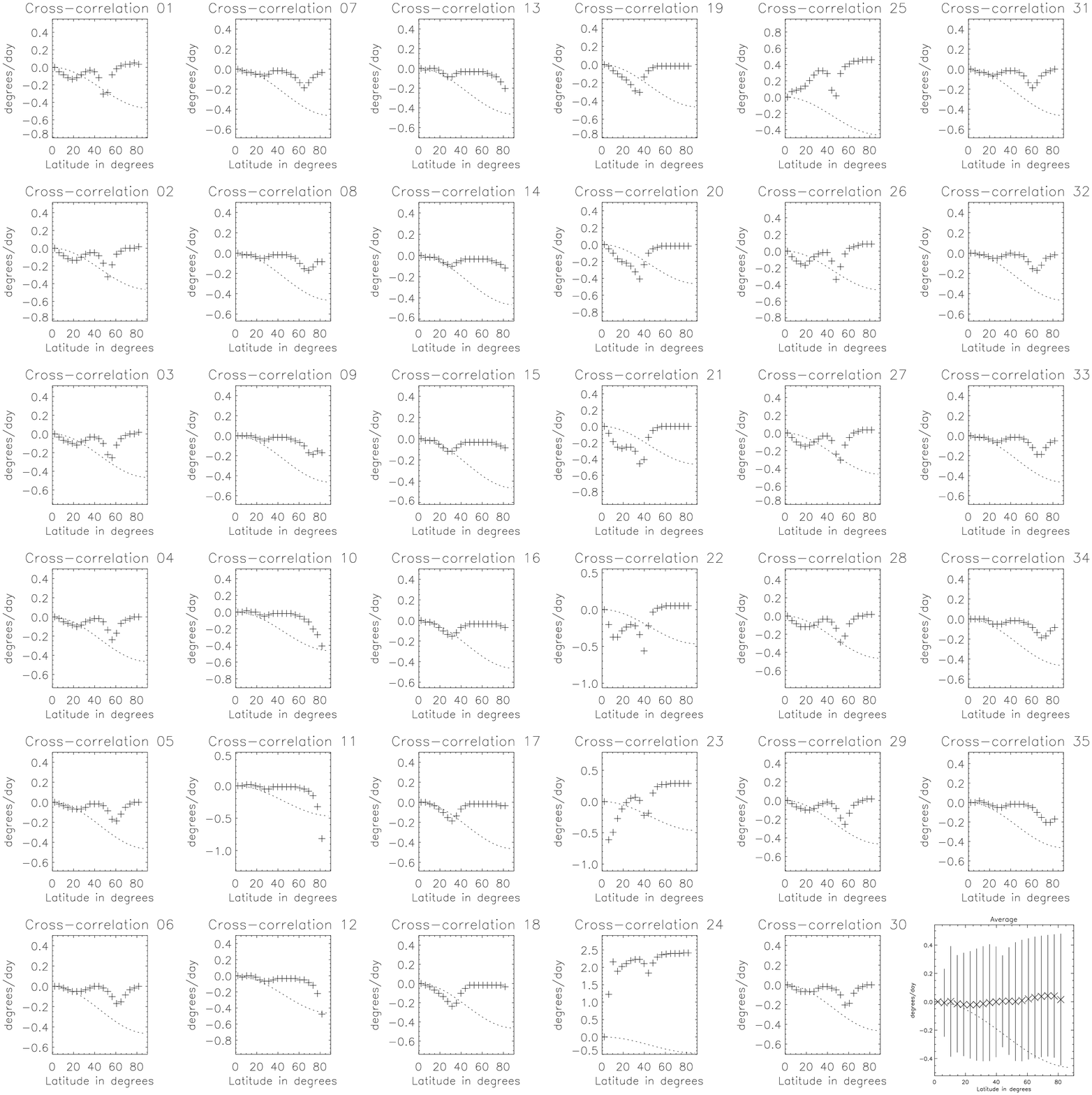}
  \caption{The results from cross-correlating the 36 snapshots.}
  \label{fig:cc}
  \end{center}
\end{figure}

The results from cross-correlating the 36 maps are shown in Fig.~\ref{fig:cc}.
The plots give the shift in degrees/day for each latitude on the 'northern' 
hemisphere. The field migration in the model is removed by normalising to the 
shift at the lowest latitude used in the investigation (2$^{\circ}$). The last 
plot in Fig.~\ref{fig:cc} shows the average of the measurements from the 
cross-correlations, with standard deviation of the measurements as the error. 
In the plots the dashed line is the input rotation at the stellar surface. The 
surface differential rotation pattern clearly changes during the activity 
cycle. It is also evident that in the measured surface differential rotation 
at the low latitudes is what one would expect from the model, but at higher 
latitudes, where most of the magnetic flux is, the correlation is very poor.

\begin{acknowledgments}
HK acknowledges the grant from IAU to help participating in the symposium.
\end{acknowledgments}

\end{document}